\documentclass[aip,jcp,reprint]{revtex4-2}
\usepackage[T1]{fontenc}
\usepackage[utf8]{inputenc}

\usepackage{amsmath}
\usepackage{amssymb}
\usepackage{calc}

\usepackage{graphicx}
\usepackage{xcolor}
\usepackage{siunitx}\DeclareSIUnit\angstrom{\text{Å}}
\usepackage[hidelinks]{hyperref}
\usepackage{algorithm}
\usepackage{algpseudocode}

\newcommand*{\vdotswithin}[1]{\mathrel{\makebox[\widthof{#1}]{\vdots}}}
\newcommand*{\odif}[1]{\mathop{}\!\mathrm{d}{#1}}
\let\Pr\undefined\DeclareMathOperator{\Pr}{\mathbf{P}}
\DeclareMathOperator{\E}{\mathbf{E}}
\newcommand*{\inner}[2]{\langle #1 , #2 \rangle}
\newcommand*{\X}[1]{X_{#1}}
\newcommand*{\bmin}{\wedge}
\newcommand*{\tran}{\mathsf{T}}
\newcommand*{\comp}{\mathsf{c}}
\newcommand*{\ind}[1]{\mathbf{1}_{#1}}
\newcommand*{\stp}[1]{T_{#1}}
\newcommand*{\stm}[1]{\tilde{T}_{#1}}

\newcommand*{\I}{\mathcal{I}}
\newcommand*{\projP}{\mathcal{P}}
\newcommand*{\projQ}{\mathcal{Q}}
\newcommand*{\A}[1]{\mathcal{A}^{#1}}
\newcommand*{\B}[1]{b^{#1}}
\newcommand*{\C}[1]{\mathcal{C}^{#1}}
\newcommand*{\ST}[2]{\mathcal{S}_{#1}^{#2}}
\newcommand*{\T}[1]{\mathcal{T}^{#1}}
\newcommand*{\dA}{\dot{\mathcal{A}}^0}
\newcommand*{\dB}{\dot{b}^0}

\newcommand*{\matC}[1]{K^{#1}}
\newcommand*{\matG}[1]{G^{#1}}
\newcommand*{\matH}[1]{h^{#1}}

\newcommand*{\soln}{u}
\newcommand*{\com}{w}
\newcommand*{\mfpt}{m}
\newcommand*{\qp}{q}
\newcommand*{\qm}{\tilde{q}}

\newcommand*{\psoln}{\hat{\soln}}
\newcommand*{\pcom}{\hat{\com}}
\newcommand*{\pmfpt}{\hat{\mfpt}}
\newcommand*{\pqp}{\hat{\qp}}
\newcommand*{\pqm}{\hat{\qm}}

\newcommand*{\gsoln}{\psoln_0}
\newcommand*{\gcom}{\pcom_0}
\newcommand*{\gmfpt}{\pmfpt_0}
\newcommand*{\gqp}{\pqp_0}
\newcommand*{\gqm}{\pqm_0}

\newcommand*{\isoln}[1]{\soln^{#1}}
\newcommand*{\icom}[1]{\com^{#1}}
\newcommand*{\imfpt}[1]{\mfpt^{#1}}
\newcommand*{\iqp}[1]{\qp^{#1}}
\newcommand*{\iqm}[1]{\qm^{#1}}

\newcommand*{\ipsoln}[1]{\psoln^{#1}}
\newcommand*{\ipcom}[1]{\pcom^{#1}}
\newcommand*{\ipmfpt}[1]{\pmfpt^{#1}}
\newcommand*{\ipqp}[1]{\pqp^{#1}}
\newcommand*{\ipqm}[1]{\pqm^{#1}}

\newcommand*{\imsoln}[2]{\delta\soln^{#1}_{#2}}
\newcommand*{\imcom}[2]{\delta\com^{#1}_{#2}}
\newcommand*{\immfpt}[2]{\delta\mfpt^{#1}_{#2}}
\newcommand*{\imqp}[2]{\delta\qp^{#1}_{#2}}
\newcommand*{\imqm}[2]{\delta\qm^{#1}_{#2}}

\newcommand*{\step}[1]{\soln_{#1}}
\newcommand*{\pstep}[1]{\psoln_{#1}}

\newcommand*{\istep}[2]{\soln^{#1}_{#2}}
\newcommand*{\ipstep}[2]{\psoln^{#1}_{#2}}

\newcommand*{\coef}{v}
\newcommand*{\basis}{\phi}

\newcommand*{\gen}{\mathcal{L}}
\newcommand*{\dt}{\epsilon}
\newcommand*{\D}[1]{\mathcal{D}({#1})}

\begin{document}

\title{Accurate estimates of dynamical statistics using memory}

\author{Chatipat Lorpaiboon}
\affiliation{Department of Chemistry and James Franck Institute, University of Chicago, Chicago, Illinois 60637, United States}
\author{Spencer C. Guo}
\affiliation{Department of Chemistry and James Franck Institute, University of Chicago, Chicago, Illinois 60637, United States}
\author{John Strahan}
\affiliation{Department of Chemistry and James Franck Institute, University of Chicago, Chicago, Illinois 60637, United States}
\author{Jonathan Weare}
\affiliation{Courant Institute of Mathematical Sciences, New York University, New York, New York 10012, United States}
\author{Aaron R. Dinner}
\email{dinner@uchicago.edu}
\affiliation{Department of Chemistry and James Franck Institute, University of Chicago, Chicago, Illinois 60637, United States}

\begin{abstract}
Many chemical reactions and molecular processes occur on timescales that are significantly longer than those accessible by direct simulation.
One successful approach to estimating dynamical statistics for such processes is to use many short time series observations of the system to construct a Markov state model (MSM), which approximates the dynamics of the system as memoryless transitions between a set of discrete states.
The dynamical Galerkin approximation (DGA) generalizes MSMs for the problem of calculating dynamical statistics, such as committors and mean first passage times, by replacing the set of discrete states with a projection onto a basis.
Because the projected dynamics are generally not memoryless, the Markov approximation can result in significant systematic error.
Inspired by quasi-Markov state models, which employ the generalized master equation to encode memory resulting from the projection, we reformulate DGA to account for memory and analyze its performance on two systems: a two-dimensional triple well and helix-to-helix transitions of the AIB\textsubscript{9} peptide.
We demonstrate that our method is robust to the choice of basis and can decrease the time series length required to obtain accurate kinetics by an order of magnitude.
\end{abstract}

\maketitle

\section{Introduction}

Models that seek to treat complex systems with high fidelity typically have many variables (e.g., the positions and velocities of all the atoms in a molecular system).  However, often the dynamics of interest of such models can be largely captured by a relatively small number of variables that are functions of the original ones (collective variables).  Projecting the dynamics to these collective variables can facilitate analysis, by both improving convergence of statistics and simplifying interpretation.  However, the coupling of the collective variables to orthogonal ones generally introduces a form of memory in which the projected dynamics depend not only on the states of the collective variables but their histories.

When the dynamics have a separation of time scales, it may be possible to find collective variables that minimize the memory, so that it can be neglected.  This is the idea behind Markov State Models (MSMs), in which the dynamics are treated as hops between discrete states, and large numbers of states are often used in the hope of minimizing the time for equilibration within states.
However, not all dynamics have a clear separation of time scales, and, even if they do, finding collective variables that minimize the memory can be challenging \cite{guttenberg_minimizing_2013}.

Thus various strategies have been introduced to account for memory.  One is to consider explicit histories of the collective variables, as in delay embedding\cite{liebert_proper_1989, das_delay-coordinate_2019, kamb2020time, thiede_galerkin_2019, strahan_long-time-scale_2021}.  However, this increases the effective number of variables, mitigating the advantages of projecting described above.  An alternative is to account for the memory through the Mori--Zwanzig formalism\cite{mori_transport_1965, zwanzig_classical_1983}.  In this approach, the projected dynamics are represented through a generalized Langevin/Master equation (GLE/GME) that involves a time integral over the collective variable histories and a kernel.  This approach is physically well-motivated and can in principle yield highly accurate dynamics.  In practice, it can be hard to determine the memory kernel.

Traditionally, researchers chose the memory kernel based on physical intuition\cite{berkowitz_memory_1981, berkowitz_generalized_1983}, and many studies still assume a specific functional form for the kernel (most often exponential) when estimating it from data.  A few studies allow for more flexibility by introducing an expansion in basis functions \cite{perico_positional_1993, kostov_mode_1997, kostov_long-time_1999,chorin_optimal_2002} or alternative fitting forms \cite{lei_data-driven_2016,grogan_data-driven_2020} for the kernel or its Laplace/Fourier transforms\cite{ayaz_non-markovian_2021}, and these show that it does not have a simple universal functional form.  However, solving for the coefficients of the expansion can be numerically ill-posed \cite{lange_collective_2006}, and data-driven parameterization of GLEs/GMEs remains an active area of research \cite{lee_multi-dimensional_2019, lin_data-driven_2021, vroylandt_likelihood-based_2022, aristoff_arbitrarily_2023, lin_regression-based_2023}.

Recently, approaching the problem from the perspective of improving the accuracy of MSMs with relatively small numbers of states\cite{cao_advantages_2020}, Huang, Markland, Montoya-Castillo and co-workers obtained the memory kernel at successive discrete time points by directly inverting a GME for the state-to-state transition matrix \cite{cao_advantages_2020, cao_integrative_2023, dominic_building_2023}.  This enabled them to significantly shorten the lag times needed to construct the transition matrix, which reduced the amount of simulation needed to converge statistics and increased the time resolution of the model; the small number of states facilitates interpretation.

Like many MSM studies, the studies described immediately above focus on computing implied time scales (via time correlation functions rather than directly from the eigenvalues of the transition matrix).  However, MSMs can be used to solve for statistics that provide more direct insight into the mechanisms of specified events, such as the committor \cite{noe_constructing_2009}, and we recently introduced a closely related framework that solves operator equations for such statistics, that we term dynamical Galerkin approximation (DGA) \cite{thiede_galerkin_2019,strahan_long-time-scale_2021}.  The goal of this paper is to show how memory can improve the accuracy of such statistics.  

To this end, we incorporate memory into DGA by interpreting it as an iterative algorithm and applying a discrete-time Mori--Zwanzig formalism.
We present formulas for computing the aforementioned statistics, and numerically demonstrate on a two-dimensional triple well and on the AIB\textsubscript{9} peptide that memory indeed improves to accuracy of these statistics.

\section{Theory}

In this section, we briefly review DGA as previously presented \cite{thiede_galerkin_2019,strahan_long-time-scale_2021}.  We then recast it as an iteration to show that its solutions can be viewed as projections; the iteration and resulting projection operators lead naturally to a discrete-time GME and, in turn, an improved estimator.  This estimator can be used without iteration.

\subsection{Form of the problem}

Let $\X{t}$ denote a time-homogeneous ergodic Markov process at time $t$.
Our goal is to compute a statistic $\soln$ of this process that satisfies an equation of the form
\begin{equation} \label{eq:exact_fix0}
    \dA \soln = -\dB,
\end{equation}
where $\dA$ is a linear operator that describes the time evolution of expectations of functions of $\X{t}$ (akin to a time derivative), and $\dB$ is the rate of change in $\soln$ at $\X{t}$.
We can interpret \eqref{eq:exact_fix0} as finding a time-invariant solution to the  equation specifying the dynamics of $\soln$,
\begin{equation} \label{eq:exact_ds0}
    \frac{\odif{\step{t}}}{\odif{t}} = \dA \step{t} + \dB.
\end{equation}

Integrating \eqref{eq:exact_ds0} over a time interval $\tau > 0$ that we call the lag time yields the Richardson iteration
\begin{equation} \label{eq:exact_ds}
    \step{t} = \step{t-\tau} + (\A{\tau} \step{t-\tau} + \B{\tau}),
\end{equation}
where $\A{\tau} = \int_0^\tau \C{\tau'} \dA \odif{\tau'} = \C{\tau} - \I$, $\B{\tau} = \int_0^\tau \C{\tau'} \dB \odif{\tau'}$, $\C{\tau} = e^{\dA \tau}$, and $\I$ is the identity operator.
Eq.~\eqref{eq:exact_ds}, as we discuss further in Section~\ref{sec:iterative}, corresponds to an iteration in $\soln$ that is consistent with the dynamics in \eqref{eq:exact_ds0}. The fixed point of the iteration is the finite time analogue of \eqref{eq:exact_fix0}:
\begin{equation} \label{eq:exact_fix}
    \A{\tau} \soln = -\B{\tau}.
\end{equation}
For example, the mean time to first enter the set $B$ starting from state $x$, $\soln(x) := \mfpt(x) = \E[\stp{B} \mid \X{0} = x]$, where $\stp{B} = \min\{t \ge 0 \mid \X{t} \in B\}$, satisfies \cite{strahan_long-time-scale_2021}
\begin{equation} \label{eq:mfpt_fix}
(\ST{B}{\tau} - \I) \mfpt = -\int_0^\tau \ST{B}{t} \ind{B^\comp} \odif{t},
\end{equation}
subject to the boundary condition $\mfpt(x) = 0$ for $x \in B$.  Here, 
\begin{equation} \label{eq:Sp}
\ST{B}{\tau} f(x) = \E[f(\X{\tau\bmin\stp{B}}) \mid \X{0} = x]
\end{equation}
is the stopped transition operator,
\begin{equation} \label{eq:mfpt_int}
\int_0^\tau \ST{B}{t} \ind{B^\comp}(x) \odif{t} = \E[\tau \bmin \stp{B} \mid \X{0} = x],
\end{equation}
$\tau\bmin\stp{B} = \min\{\tau,\stp{B}\}$, and $\ind{B^\comp}(x)$ is an indicator function that equals one for $x \in B^\comp$ and zero for $x \notin B^\comp$.  We define other statistics of interest in Section \ref{sec:statistics}.

\subsection{Dynamical Galerkin approximation (DGA)}

DGA solves \eqref{eq:exact_fix} by expanding $\soln$ in a basis and estimating the resulting matrix elements by averages over random samples.
We represent the basis by a vector $\basis$ of functions that satisfy homogeneous boundary conditions.
Then we approximate $\soln$ by 
\begin{equation}
\label{eq:dga_expand}
\psoln = \gsoln + \basis^\tran \coef,
\end{equation}
where $\gsoln$ is a ``guess function'' for $\soln$ that satisfies the boundary conditions, $\coef$ is a vector of coefficients, and we use the symbol $\hat{\ }$ to denote functions that can be represented by \eqref{eq:dga_expand}.
To solve for the coefficients, we substitute \eqref{eq:dga_expand} into \eqref{eq:exact_fix}, multiply from the left by $\basis$, and integrate over an arbitrary distribution of samples of initial conditions $\mu$.  This yields
\begin{equation} \label{eq:dga}
\matG{\tau} \coef = -\matH{\tau},
\end{equation}
where 
\begin{align}
\matG{\tau} & = \inner{\basis}{\A{\tau} \basis^\tran}, \\
\matH{\tau} & = \inner{\basis}{\A{\tau} \gsoln + \B{\tau}},
\end{align}
and $\inner{f}{g} = \int f(x) g(x) \mu(x) \odif{x}$.
The matrix $\matG{\tau}$ and vector $\matH{\tau}$ are estimated from averages over trajectories of length $\tau$; explicit expressions are given in Sec.~\ref{sec:statistics}.
Given these estimates, \eqref{eq:dga} can be solved directly for $\coef$.

\subsection{\label{sec:iterative} Iterative solution}

The form in \eqref{eq:exact_ds} and our recent use of it in the context of inexact numerical linear algebra \cite{strahan_inexact_2023} suggests that one can view DGA as the fixed point of the \textit{inexact} iteration
\begin{subequations} \label{eq:inexact}
\begin{align}
    \istep{\tau}{t} & = \ipstep{\tau}{t-\tau} + (\A{\tau} \ipstep{\tau}{t-\tau} + \B{\tau}),
    \label{eq:inexact1} \\
    \ipstep{\tau}{t} & = \ipstep{\tau}{t-\tau} + \projP (\istep{\tau}{t} - \ipstep{\tau}{t-\tau}),
    \label{eq:inexact2}
\end{align}
\end{subequations}
where $\projP$ is an operator that projects functions onto linear combinations of the basis functions:
\begin{equation}
\projP f = \basis^\tran (\matC{0})^{-1} \inner{\basis}{f},
\end{equation}
and
\begin{equation}
\matC{t} = \inner{\basis}{\C{t} \basis^\tran}.
\end{equation}
We use the superscript $\tau$ on $\istep{\tau}{t}$ to distinguish that it comes from the inexact iteration \eqref{eq:inexact} involving the projected function $\ipstep{\tau}{t}$ (and thus a layer of approximation) rather than the exact iteration \eqref{eq:exact_ds}.
Fixed points are indicated without subscripts: $\isoln{\tau} = \lim_{t \to \infty} \istep{\tau}{t}$ and $\ipsoln{\tau} = \lim_{t \to \infty} \ipstep{\tau}{t}$.

We now identify the approximation associated with DGA by comparing the inexact iteration \eqref{eq:inexact} with the \textit{exact} iteration
\begin{subequations} \label{eq:exact}
\begin{align}
\step{t} & = \step{t-\tau} + (\A{\tau} \step{t-\tau} + \B{\tau}),
\label{eq:exact1} \\
\pstep{t} & = \pstep{t-\tau} + \projP (\step{t} - \pstep{t-\tau}).
\label{eq:exact2}
\end{align}
\end{subequations}
Let us define the complementary projection
\begin{equation}
\projQ = \I - \projP.
\end{equation}
We note that $\projQ (\pstep{t}-\pstep{t'}) = 0$ for any $t$ and $t'$ since $\projQ \basis = 0$ and $\pstep{t}-\pstep{t'} = \basis^\tran \coef$ for some $v$ (we work with differences because we do not assume $\projQ \gsoln=0$).
By substituting $\projP = \I - \projQ$ into \eqref{eq:exact2} and replacing $t$ with $t-\tau$, we find that $\step{t-\tau} = \pstep{t-\tau} + \projQ (\step{t-\tau} - \pstep{t-2\tau})$; in turn, by substituting this expression into \eqref{eq:exact1}, we can express \eqref{eq:exact} as
\begin{subequations} \label{eq:exactp}
\begin{align}
    \step{t} & = \pstep{t-\tau} + (\A{\tau} \pstep{t-\tau} + \B{\tau}) + \C{\tau} \projQ (\step{t-\tau} - \pstep{t-2\tau}), \label{eq:exactp1} \\
    \pstep{t} & = \pstep{t-\tau} + \projP (\step{t} - \pstep{t-\tau}). \label{eq:exactp2}
\end{align}
\end{subequations}
The fixed point $\psoln = \lim_{t \to \infty} \pstep{t}$ satisfies $\projP (\A{\tau} \psoln + \B{\tau}) + \projP \C{\tau} \projQ (\soln - \psoln) = 0$.
Comparing \eqref{eq:inexact1} and \eqref{eq:exact1} shows that they both use the previous estimate ($\ipstep{\tau}{t-\tau}$ or $\istep{\tau}{t-\tau}$, respectively) to construct an improved estimate $\istep{\tau}{t}$.
Because $\ipstep{\tau}{t-\tau}$ can be represented using the basis, we only need length $\tau$ trajectories to approximate $\istep{\tau}{t}$.
Hence, DGA makes the approximation $\projP \C{\tau} \projQ (\soln - \psoln) \approx 0$, which avoids the need for trajectories longer than $\tau$.
In contrast, the exact iteration requires infinite-length trajectories since the last term of \eqref{eq:exactp1} prevents $\pstep{t}$ from depending on $\pstep{t-\tau}$ alone.

To understand the need for memory, we observe that \eqref{eq:exact1} specifies Markovian dynamics for $\step{t}$: the value of $\step{t}$ at current iteration $t$ is fully determined by the value of $\step{t-\tau}$ at the last iteration $t-\tau$.  DGA amounts to making the approximation that the dynamics of the \emph{projected} function $\pstep{t}$ observed at discrete time intervals $\tau$ is Markovian.
This perspective suggests that we can derive an improved estimator by mitigating the Markov approximation.
We do this by using projected statistics at multiple times (i.e., using memory), rather than a single time,
as we now describe.

\subsection{Dynamical Galerkin approximation with memory}

Our starting point is \eqref{eq:exactp1} with $\tau$ replaced by $\sigma$, where $\tau/\sigma$ is a positive integer:
\begin{equation} \label{eq:iter}
\step{t} = \pstep{t-\sigma} + (\A{\sigma} \pstep{t-\sigma} + \B{\sigma}) + \C{\sigma} \projQ (\underline{\step{t-\sigma}} - \pstep{t-2\sigma}).
\end{equation}
The underlined term is not representable by the basis and prevents solving for $\psoln$ by fixed point iteration with length $\tau$ trajectories.
By repeatedly substituting \eqref{eq:iter}, we obtain
\begin{alignat}{2}
\step{t}
    & = \pstep{t-\sigma}
        && + \sum_{n=1}^2 (\C{\sigma} \projQ)^{n-1} (\A{\sigma} \pstep{t-n\sigma} + \B{\sigma}) \nonumber \\
        &&& + (\C{\sigma} \projQ)^2 (\underline{\step{t-2\sigma}} - \pstep{t-3\sigma}) \nonumber \\
    & \vdotswithin{=} \nonumber \\
    & = \pstep{t-\sigma}
        && + \sum_{n=1}^{\tau/\sigma} (\C{\sigma} \projQ)^{n-1} (\A{\sigma} \pstep{t-n\sigma} + \B{\sigma}) \nonumber \\
        &&& + (\C{\sigma} \projQ)^{\tau/\sigma} (\underline{\step{t-\tau}} - \pstep{t-\tau-\sigma}),
\end{alignat}
where the underline indicates the location of the next substitution.
We now have an alternative iteration to \eqref{eq:exactp} that depends on the last $\tau/\sigma$ projections:
\begin{subequations} \label{eq:exactm}
\begin{alignat}{2}
\step{t} & = \pstep{t-\sigma}
    && + \sum_{n=1}^{\tau/\sigma}
        (\C{\sigma} \projQ)^{n-1}
        (\A{\sigma} \pstep{t-n\sigma} + \B{\sigma}) \nonumber \\
    &&& + (\C{\sigma} \projQ)^{\tau/\sigma} (\step{t-\tau} - \pstep{t-\tau-\sigma}),
\label{eq:exactm1} \\
\pstep{t} & = \pstep{t-\sigma} && + \projP (\step{t} - \pstep{t-\sigma}).
\label{eq:exactm2}
\end{alignat}
\end{subequations}
Equation \eqref{eq:exactm1} is similar to a discrete-time GME (e.g., Ref.~\onlinecite{cao_advantages_2020}) or a discrete-time Mori--Zwanzig decomposition (e.g., Ref.~\onlinecite{darve_computing_2009} for eigendecomposition of the transition operator).
It describes $\step{t}$ at iteration $t$ by its projection $\pstep{t-\sigma}$ at iteration $t-\sigma$ (\emph{Markov approximation}, first term and $n=1$ of the second term), the projections $\pstep{t-n\sigma}$ at previous iterations $t-n\sigma$ (\emph{memory}, $n>1$ of the second term), and the complementary projection $\projQ \step{t-\tau}$ at time $t-\tau$ (\emph{orthogonal dynamics}, last term).
The approximation used in DGA is the first term of \eqref{eq:exactm1} (with $\tau$ replacing $\sigma$).
The remaining terms correct for the Markov approximation by introducing memory.

Omitting the last term of \eqref{eq:exactm1} yields a computable iteration
\begin{subequations} \label{eq:inexactm}
\begin{align}
\label{eq:inexactm1}
\istep{\sigma,\tau}{t} & = \ipstep{\sigma,\tau}{t-\sigma} + \sum_{n=1}^{\tau/\sigma} (\C{\sigma} \projQ)^{n-1} (\A{\sigma} \ipstep{\sigma,\tau}{t-n\sigma} + \B{\sigma}),
\\
\label{eq:inexactm2}
\ipstep{\sigma,\tau}{t} & = \ipstep{\sigma,\tau}{t-\sigma} + \projP (\istep{\sigma,\tau}{t} - \ipstep{\sigma,\tau}{t-\sigma}),
\end{align}
\end{subequations}
with a fixed point that solves
\begin{equation}  \label{eq:inexactm2_fix}
    \sum_{n=1}^{\tau/\sigma}
        \projP (\C{\sigma} \projQ)^{n-1}
        (\A{\sigma} \ipsoln{\sigma,\tau} + \B{\sigma})
    = 0.
\end{equation}

Comparing \eqref{eq:exactm} and \eqref{eq:inexactm2_fix}, we see that the latter corresponds to making the approximation $\projP (\C{\sigma} \projQ)^{\tau/\sigma} (\soln - \psoln) \approx 0$.
When the basis captures much of the slower dynamics of the system, we expect the DGA with memory approximation $\projP (\C{\sigma} \projQ)^{\tau/\sigma} (\soln - \psoln) \approx 0$ to be better than the DGA approximation $\projP \C{\tau} \projQ (\soln - \psoln) \approx 0$ because the additional applications of $\projQ$ remove these slower dynamics. We show this numerically in Sec.~\ref{sec:examples}.

\subsection{Algorithm}

\begin{figure}[tb]
\begin{algorithm}[H]
    \caption{\label{fig:algorithm}
        DGA with memory
    }
    \begin{algorithmic}
        \For{$n \in \{1,\ldots,\tau/\sigma\}$}
            \State $\matC{n\sigma} = \inner{\basis}{\C{n\sigma} \basis^\tran}$
            \State $\matG{n\sigma} = \inner{\basis}{\A{n\sigma} \basis^\tran}$
            \State $\matH{n\sigma} = \inner{\basis}{\A{n\sigma} \gsoln + \B{n\sigma}}$
            \State $\matG{\sigma,n\sigma} = \matG{n\sigma} - \sum_{n'=1}^{n-1} \matC{(n-n')\sigma} (\matC{0})^{-1} \matG{\sigma,n'\sigma}$
            \State $\matH{\sigma,n\sigma} = \matH{n\sigma} - \sum_{n'=1}^{n-1} \matC{(n-n')\sigma} (\matC{0})^{-1} \matH{\sigma,n'\sigma}$
        \EndFor
        \State $\coef = -(\matG{\sigma,\tau})^{-1} \matH{\sigma,\tau}$
        \State $\ipsoln{\sigma,\tau} = \gsoln + \basis^\tran \coef$
        \State $\begin{aligned}
                \textstyle \isoln{\sigma,\tau}
                & \textstyle = \ipsoln{\sigma,\tau} + (\A{\tau} \ipsoln{\sigma,\tau} + \B{\tau}) \\
                & \textstyle \mathrel{\phantom{=}} {} - \sum_{n=1}^{\tau/\sigma} \C{\tau-n\sigma} \basis^\tran (\matC{0})^{-1} (\matG{\sigma,n\sigma} \coef + \matH{\sigma,n\sigma})
                \end{aligned}$
    \end{algorithmic}
\end{algorithm}
\end{figure}

In this section, we derive and summarize the algorithm (Algorithm~\ref{fig:algorithm}).  As in \eqref{eq:dga_expand}, we expand $\ipsoln{\sigma,\tau}$ in a basis.
Substituting $\ipsoln{\sigma,\tau} = \gsoln + \basis^\tran \coef$ into \eqref{eq:inexactm2_fix} and applying $\inner{\basis_i}{\projP f} = \inner{\basis_i}{f}$ leads to the linear system
\begin{equation} \label{eq:dga_mem}
\matG{\sigma,\tau} \coef = -\matH{\sigma,\tau},
\end{equation}
where 
\begin{align}
\matG{\sigma,\tau} & = \sum_{n=1}^{\tau/\sigma} \inner{\basis}{(\C{\sigma} \projQ)^{n-1} \A{\sigma} \basis^\tran}, \\
\matH{\sigma,\tau} & = \sum_{n=1}^{\tau/\sigma} \inner{\basis}{(\C{\sigma} \projQ)^{n-1} (\A{\sigma} \gsoln + \B{\sigma})}.
\end{align}

We can calculate matrices $\matG{\sigma,\tau}$ and $\matH{\sigma,\tau}$ by expanding each $\projQ$ in the expectations as $\inner{f}{\projQ g} = \inner{f}{g}-\inner{f}{\projP g}=\inner{f}{g} - \inner{f}{\basis^\tran} (\matC{0})^{-1} \inner{\basis}{g}$ until none remain.
For instance, to calculate the term $\matH{\sigma,3\sigma}$, we expand:
\begin{align*}
\matH{\sigma,3\sigma}
    & = \inner{\basis}{\B{\sigma} + \C{\sigma} \projQ \B{\sigma} + (\C{\sigma} \projQ)^2 \B{\sigma}} \\
    & =
        \inner{\basis}{\B{\sigma} + \C{\sigma} \B{\sigma} + \C{2\sigma} \B{\sigma}} \\
    & \mathrel{\phantom{=}}
        - \inner{\basis}{\C{\sigma} \projP (\B{\sigma} + \C{\sigma} \projQ \B{\sigma})}
        - \inner{\basis}{\C{2\sigma} \projP \B{\sigma}} \\
    & =
        \matH{3\sigma}
        - \matC{\sigma} (\matC{0})^{-1} \underline{\matH{\sigma,2\sigma}}
        - \matC{2\sigma} (\matC{0})^{-1} \matH{\sigma}, \\
\matH{\sigma,2\sigma}
    & = \inner{\basis}{\B{\sigma} + \C{\sigma} \projQ \B{\sigma}} \\
    & = \inner{\basis}{\B{\sigma} + \C{\sigma} \B{\sigma}} - \inner{\basis}{\C{\sigma} \projP \B{\sigma}} \\
    & = \matH{2\sigma} - \matC{\sigma} (\matC{0})^{-1} \matH{\sigma}.
\end{align*}
The underlined expectations contain $\projQ$ and must be expanded in terms of matrices that do not contain $\projQ$, which can be directly evaluated as averages over random samples.
We have rearranged and grouped terms for numerical stability because $\matG{\sigma,\tau}$ and $\matH{\sigma,\tau}$ can be very sensitive to how they are computed.
Using this approach, we can derive the iterative formulas
\begin{subequations} \label{eq:invert}
\begin{align}
\matG{\sigma,\tau} & = \matG{\tau} - \sum_{n=1}^{(\tau/\sigma)-1} \matC{\tau-n\sigma} (\matC{0})^{-1} \matG{\sigma,n\sigma},
\\
\matH{\sigma,\tau} & = \matH{\tau} - \sum_{n=1}^{(\tau/\sigma)-1} \matC{\tau-n\sigma} (\matC{0})^{-1} \matH{\sigma,n\sigma}.
\end{align}
\end{subequations}
This is similar to the procedure in Ref.~\onlinecite{cao_advantages_2020}, where the discrete-time GME is directly inverted.

After we solve \eqref{eq:dga_mem}, we can compute a stochastic approximation of $\soln$:
\begin{equation} \label{eq:inexactm1_fix}
\isoln{\sigma,\tau} =
    \ipsoln{\sigma,\tau}
    + \sum_{n=1}^{\tau/\sigma}
        (\C{\sigma} \projQ)^{n-1}
        (\A{\sigma} \ipsoln{\sigma,\tau} + \B{\sigma}),
\end{equation}
which is the fixed point of \eqref{eq:inexactm1}.
In this case, we expand each $\projQ f = f - \basis^\tran (\matC{0})^{-1} \inner{\basis}{f}$ from left to right until only $\projQ$s inside expectations remain (which can be grouped into matrices $\matG{\sigma,n\sigma}$ and $\matH{\sigma,n\sigma}$):
\begin{equation} \label{eq:isoln_invert}
\isoln{\sigma,\tau} =
    \ipsoln{\sigma,\tau} + (\A{\tau} \ipsoln{\sigma,\tau} + \B{\tau})
    - \sum_{n=1}^{\tau/\sigma} \C{\tau-n\sigma} \imsoln{\sigma,\tau}{n},
\end{equation}
where $\imsoln{\sigma,\tau}{n} = \basis^\tran (\matC{0})^{-1} (\matG{\sigma,n\sigma} \coef + \matH{\sigma,n\sigma})$.
As in \eqref{eq:invert}, we have combined terms in \eqref{eq:isoln_invert} for numerical stability.

\section{Dynamical statistics \label{sec:statistics}}

In this section, we introduce the statistics that we want to compute and the equations that they solve.
We describe a variety of statistics beyond the MFPT that can be solved using DGA or DGA with memory.
Provided that a statistic can be written in the form \eqref{eq:exact_fix}, we can solve for it with DGA or DGA with memory by evaluating \eqref{eq:dga} or \eqref{eq:dga_mem}, respectively, with appropriate $\matC{t}$, $\matG{t}$, and $\matH{t}$ matrices.

\subsection{Stationary distribution}

The stationary distribution $\pi$ is invariant with respect to time translation and therefore
\begin{equation}
    \int \pi(x) f(x) \odif{x} = \int \pi(x) \T{t} f(x) \odif{x}
\end{equation}
for all functions $f$, where the transition operator
\begin{equation}
    \T{t} f(x) = \E[f(\X{t}) \mid \X{0} = x]
\end{equation}
propagates expectations of functions $f$ forward-in-time.
The sampling distribution $\mu$ can be reweighted to obtain the stationary distribution $\pi$ via $\com = \pi / \mu$ \cite{wu_variational_2017,strahan_long-time-scale_2021}, which satisfy an operator equation in the form of \eqref{eq:exact_fix}:
\begin{equation} \label{eq:pi_bvp}
((\T{t})^\dagger - \I) \com = 0,
\end{equation}
where $(\T{t})^\dagger$ denotes the adjoint of $\T{t}$ with respect to $\mu$: $\inner{f}{(\T{t})^\dagger g} = \inner{g}{\T{t} f}$ for all functions $f$ and $g$.
We expand $\pcom = \gcom + \basis^\tran \coef$, where $\int \gcom(x) \mu(x) \odif{x} = 1$ and $\int \basis(x) \mu(x) \odif{x} = 0$;
these conditions ensure that the resulting estimates of $\pcom$ and $\com$ are normalized.
The corresponding matrices are
\begin{subequations} \label{eq:pi_mats}
\begin{align}
    \matC{t}
        & = \inner{\basis}{(\T{t})^\dagger \basis^\tran} \nonumber \\
        & = \E[\basis(\X{t}) \basis^\tran(\X{0}) \mid \X{0} \sim \mu], \label{eq:CtL} \\
    \matG{t}
        & = \inner{\basis}{((\T{t})^\dagger - \I) \basis^\tran} \nonumber \\
        & = \E[(\basis(\X{t}) - \basis(\X{0})) \basis^\tran(\X{0}) \mid \X{0} \sim \mu], \\
    \matH{t}
        & = \inner{\basis}{((\T{t})^\dagger - \I) \gcom} \nonumber \\
        & = \E[(\basis(\X{t}) - \basis(\X{0})) \gcom(\X{0}) \mid \X{0} \sim \mu],
\end{align}
\end{subequations}
and we approximate expectations of functions $f$ with respect to $\pi$ as
\begin{equation} \label{eq:w_stoc}
\begin{aligned}
    \inner{f}{\icom{\sigma,\tau}}
    = \E \biggl[
            & f(\X{\tau}) \ipcom{\sigma,\tau}(\X{0}) \\
            & - \sum_{n=1}^{\tau/\sigma} f(\X{\tau-n\sigma}) \imcom{\sigma,\tau}{n}(\X{0})
        \biggm| \X{0} \sim \mu \biggr].
\end{aligned}
\end{equation}

\subsection{Mean first passage time}

The mean first passage time (MFPT)
\begin{equation}
\mfpt(x) = \E[\stp{B} \mid \X{0} = x]
\end{equation}
is the expected time until a trajectory starting at $x$ first enters $B$.  As noted above, it satisfies an equation of the form \eqref{eq:exact_fix}:
\begin{equation} \label{eq:mfpt_bvp}
(\ST{B}{t} - \I) \mfpt = -\int_0^t \ST{B}{t'} \ind{B^\comp} \odif{t'},
\end{equation}
subject to the boundary condition $\mfpt(x) = 0$ for $x \in B$.
Physically, this equation says that each application of $\ST{B}{t}$ reduces the mean remaining time to exit by $\int_0^t \ST{B}{t'} \ind{B^\comp} \odif{t'} = \E[t\bmin\stp{B} \mid \X{0} = x]$.
The corresponding matrices are
\begin{subequations} \label{eq:mfpt_mats}
\begin{align}
\matC{t}
    & = \inner{\basis}{\ST{B}{t} \basis^\tran} \nonumber \\
    & = \E[\basis(\X{0}) \basis^\tran(\X{t\bmin\stp{B}}) \mid \X{0} \sim \mu], \\
\matG{t}
    & = \inner{\basis}{(\ST{B}{t} - \I) \basis^\tran} \nonumber \\
    & = \E[\basis(\X{0}) (\basis^\tran(\X{t\bmin\stp{B}}) - \basis^\tran(\X{0})) \mid \X{0} \sim \mu], \\
\matH{t}
    & = \biggl\langle\basis,(\ST{B}{t} - \I) \gmfpt + \int_0^t \ST{B}{t'} \ind{B^\comp} \odif{t'}\biggr\rangle \nonumber \\
    & \begin{aligned}[b] {} = \E[\basis(\X{0}) (& \gmfpt(\X{t\bmin\stp{B}}) - \gmfpt(\X{0}) \\ & + (t\bmin\stp{B})) \mid \X{0} \sim \mu], \end{aligned} \label{eq:mfpt_h}
\end{align}
\end{subequations}
and $\mfpt$ can be approximated from its projection using
\begin{equation}
\begin{aligned}[b]
\imfpt{\sigma,\tau}(x) = \E \biggl[
    & \ipmfpt{\sigma,\tau}(\X{\tau\bmin\stp{B}}) + (\tau\bmin\stp{B}) \\
    & - \sum_{n=1}^{\tau/\sigma}
        \immfpt{\sigma,\tau}{n}(\X{(\tau-n\sigma)\bmin\stp{B}})
    \biggm| \X{0} = x \biggr].
\end{aligned}
\end{equation}

\subsection{Forward committor}

The forward committor
\begin{equation}
\qp(x) = \E[\ind{B}(\X{\stp{A \cup B}}) \mid \X{0} = x]
\end{equation}
is the probability that a trajectory starting at $x$ will enter $B$ before $A$.
The forward committor satisfies
\begin{equation} \label{eq:qp_bvp}
(\ST{A \cup B}{t} - \I) \qp = 0
\end{equation}
with boundary condition $\qp(x) = \ind{B}(x)$ for $x \in A \cup B$.
The corresponding matrices are
\begin{subequations} \label{eq:qp_mats}
\begin{align}
\matC{t}
    & = \inner{\basis}{\ST{A \cup B}{t} \basis^\tran} \nonumber \\
    & = \E[\basis(\X{0}) \basis(\X{t\bmin\stp{A \cup B}}) \mid \X{0} \sim \mu], \\
\matG{t}
    & = \inner{\basis}{(\ST{A \cup B}{t} - \I) \basis^\tran} \nonumber \\
    & = \E[\basis(\X{0}) (\basis(\X{t\bmin\stp{A \cup B}}) - \basis(\X{0})) \mid \X{0} \sim \mu], \\
\matH{t}
    & = \inner{\basis}{(\ST{A \cup B}{t} - \I) \gqp} \nonumber \\
    & = \E[\basis(\X{0}) (\gqp(\X{t\bmin\stp{A \cup B}}) - \gqp(\X{0})) \mid \X{0} \sim \mu],
\end{align}
\end{subequations}
and $\qp$ can be approximated as
\begin{equation}
\begin{aligned}[b]
\iqp{\sigma,\tau}(x) = \E \biggl[
    & \ipqp{\sigma,\tau}(\X{\tau\bmin\stp{A \cup B}}) \\
    & - \sum_{n=1}^{\tau/\sigma}
        \imqp{\sigma,\tau}{n}(\X{(\tau-n\sigma)\bmin\stp{A \cup B}})
    \biggm| \X{0} = x \biggr].
\end{aligned}
\end{equation}

\subsection{Backward committor}

The backward committor is the probability that a trajectory ending at $x$ exited $A$ after $B$,
\begin{equation}
    \qm(x) = \E[\ind{A}(\X{-\stm{A \cup B}}) \mid \X{0} = x],
\end{equation}
where $\stm{A \cup B} = \min\{t \ge 0 \mid \X{-t} \in A \cup B\}$.
The backward committor satisfies
\begin{equation} \label{eq:qm_bvp}
(\ST{A \cup B}{-t} - \I) \qm = 0
\end{equation}
with boundary condition $\qm(x) = \ind{A}(x)$ for $x \in A \cup B$. \begin{equation} \label{eq:Sm}
\ST{A \cup B}{-t} f(x) = \E[f(\X{-(t\bmin\stm{A \cup B})}) \mid \X{0} = x]
\end{equation}
is the stopped transition operator for the time-reversed process.
We take conditional expectations backward-in-time with respect to the \emph{time-reversed process}: given an infinite, statistically stationary, trajectory $X$, we look backward in time from each time $t$ where $\X{t} = x$.
Mathematically, 
\begin{align}
    & \E[f(\X{-(t\bmin\stm{A \cup B})}) \mid \X{0} = x] \nonumber \\
    & \quad {} = \int \E[f(\X{-(t\bmin\stm{A \cup B})}) \mid \X{0} = x, \X{-\tau} = x'] \pi(x') \odif{x'},
\end{align}
which follows from the definition of conditional probability and $\pi(x) = \Pr[\X{-\tau} = x]$. Because $\pi(x) = \com(x) \mu(x) = \Pr[\X{0} = x] = \Pr[\X{-\tau} = x]$, 
\begin{align}
    & \inner{g}{\ST{A \cup B}{-t} f} \nonumber \\
    & \quad {} = \int \E[g(\X{0}) f(\X{-(t\bmin\stm{A \cup B})}) \mid \X{0} = x] \mu(x) \odif{x} \nonumber \\
    & \quad {} = \E \biggl[ g(\X{0}) f(\X{-(t\bmin\stm{A \cup B})}) \frac{\com(\X{-\tau})}{\com(\X{0})} \biggm| \X{-\tau} \sim \mu \biggr],
    \label{eq:g_Sm_f}
\end{align}
as we show explicitly in Ref.~\onlinecite{strahan_long-time-scale_2021} (with the zero of time shifted).
The corresponding matrices are
\begin{subequations} \label{eq:qm_mats}
\begin{align}
\matC{t}
    & = \inner{\basis}{\ST{A \cup B}{-t} \basis^\tran} \nonumber \\
    & = \E\biggl[\basis(\X{0}) \basis(\X{-(t\bmin\stm{A \cup B})}) \frac{\com(\X{-\tau})}{\com(\X{0})} \biggm| \X{-\tau} \sim \mu\biggr], \\
\matG{t}
    & = \inner{\basis}{(\ST{A \cup B}{-t} - \I) \basis^\tran} \nonumber \\
    & \begin{aligned}[b]
        {} = \E\biggl[
        & \basis(\X{0}) (\basis(\X{-(t\bmin\stm{A \cup B})}) - \basis(\X{0})) \\
        & \times \frac{\com(\X{-\tau})}{\com(\X{0})} \biggm| \X{-\tau} \sim \mu\biggr],
        \end{aligned} \\
\matH{t}
    & = \inner{\basis}{(\ST{A \cup B}{-t} - \I) \gqm} \nonumber \\
    & \begin{aligned}[b]
        {} = \E\biggl[
        & \basis(\X{0}) (\gqm(\X{-(t\bmin\stm{A \cup B})}) - \gqm(\X{0})) \\
        & \times \frac{\com(\X{-\tau})}{\com(\X{0})} \biggm| \X{-\tau} \sim \mu\biggr].
        \end{aligned}
\end{align}
\end{subequations}
We estimate $\qm$ as
\begin{equation}
\begin{aligned}[b]
\iqm{\sigma,\tau}(x) = \E \biggl[
    & \ipqm{\sigma,\tau}(\X{-(\tau\bmin\stm{A \cup B})}) \\
    & - \sum_{n=1}^{\tau/\sigma} \imqm{\sigma,\tau}{n}(\X{-((\tau-n\sigma)\bmin\stm{A \cup B})})
    \biggm| \X{0} = x \biggr].
\end{aligned}
\end{equation}

\section{\label{sec:examples} Numerical examples}

In this section, we demonstrate our method on three systems.
We first illustrate our method with a two-dimensional triple-well potential, for which an exact reference can be calculated using finite-differences.
We next test our method on a more complex system, AIB\textsubscript{9}, for which we have long, unbiased trajectories as a reference.

\subsection{\label{sec:threehole} Two-dimensional triple-well}

\begin{figure}[tb]
    \includegraphics{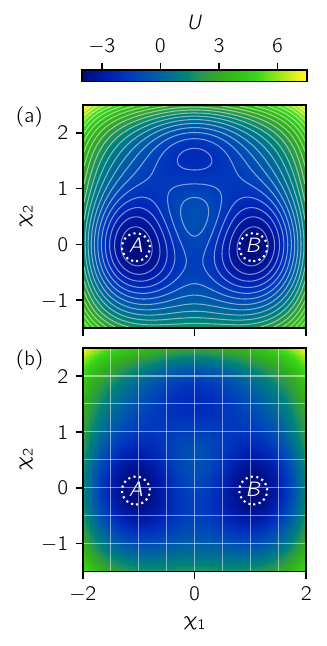}
    \caption{\label{fig:three_pmf}
        Potential energy surface of the two-dimensional triple-well.
        The boundaries of sets $A$ and $B$ are indicated using dotted lines.
        (a) Contours are drawn every $0.5 \beta^{-1}$.
        (b) $8 \times 8$ grid basis used for DGA.
    }
\end{figure}

\begin{figure*}[tb]
    \includegraphics{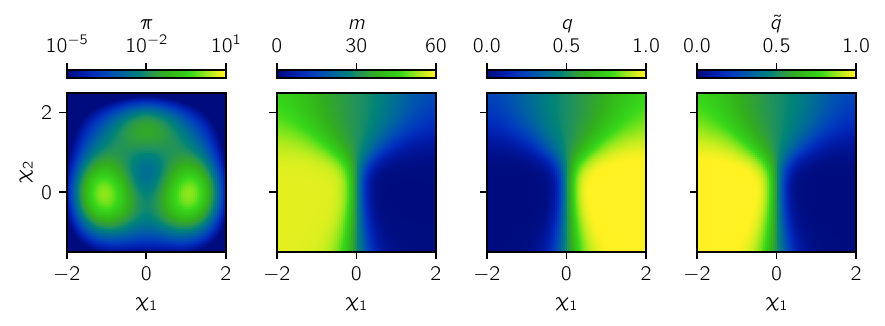}
    \caption{\label{fig:threehole_ref}
        Reference statistics for the two-dimensional triple-well.
    }
\end{figure*}

\begin{figure*}[tb]
    \includegraphics{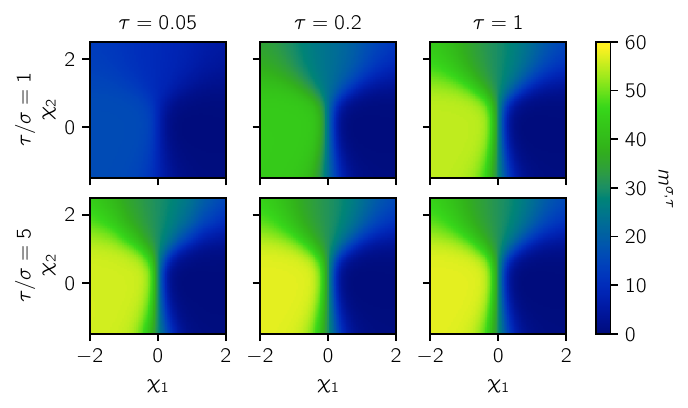}
    \caption{\label{fig:threehole_mfpt_compare}
        Mean first passage time for the two-dimensional triple-well, estimated using DGA ($\tau/\sigma=1$) and DGA with memory ($\tau/\sigma=5$).
    }
\end{figure*}

\begin{figure*}[tb]
    \includegraphics{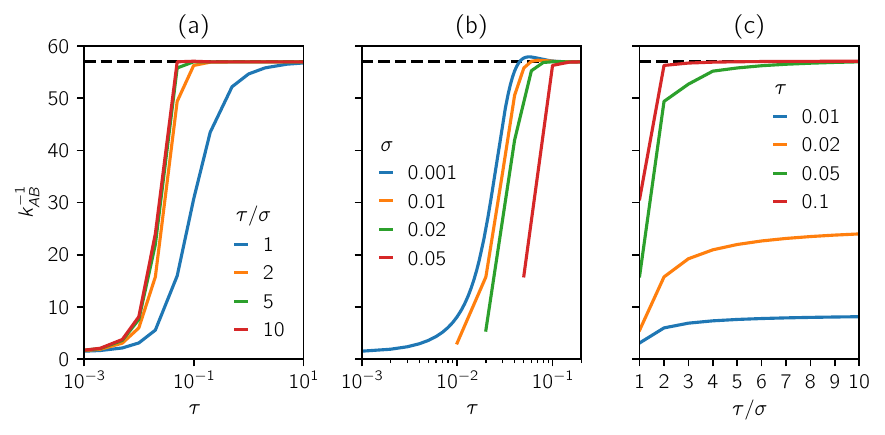}
    \caption{\label{fig:threehole_mfpt_error}
        Dependence of the estimated inverse rate constant $k_{AB}^{-1}$ for the two-dimensional triple-well on (a) $\tau$ with a fixed number of memory terms, (b) $\tau$ with fixed $\sigma$, and (c) the number of memory terms with fixed $\tau$. The true value of the inverse rate constant is indicated by the dashed line.
    }
\end{figure*}

\begin{figure*}[tb]
    \includegraphics{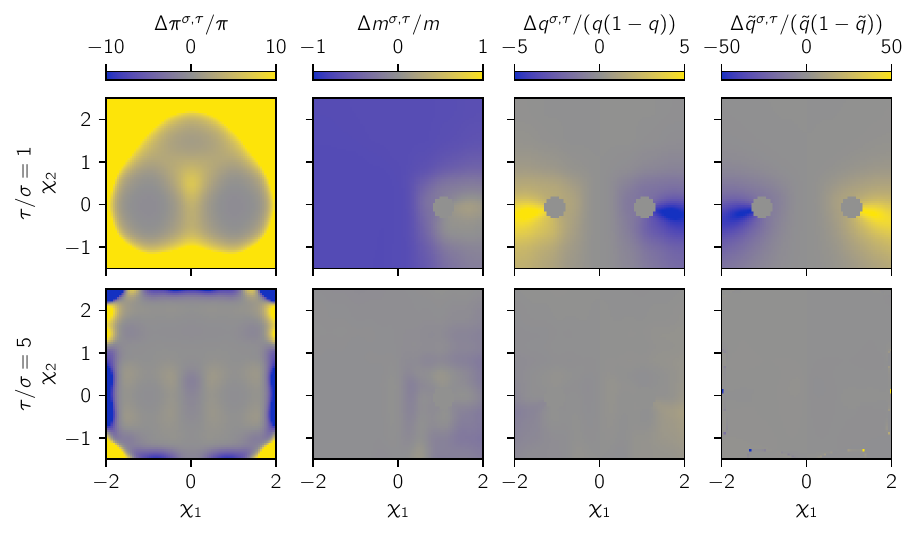}
    \caption{\label{fig:threehole_pw_rel_error}
        Relative error in the DGA ($\tau/\sigma=1$) and DGA with memory ($\tau/\sigma=5$) estimates for each dynamical statistic at $\tau=0.05$, for the two-dimensional triple-well.
    }
\end{figure*}
    
\begin{figure*}[tb]
    \includegraphics{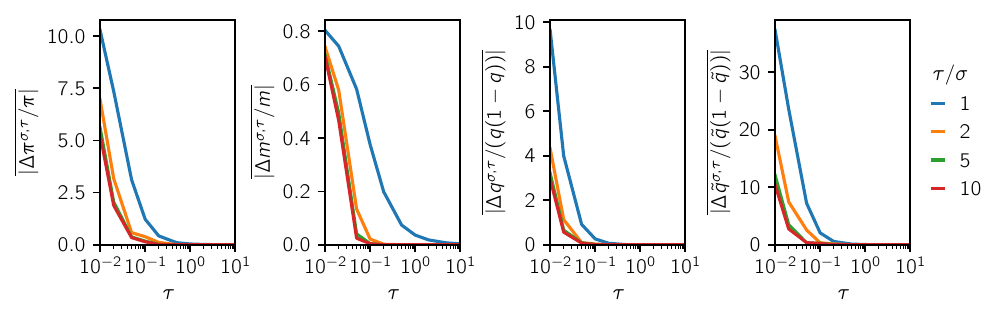}
    \caption{\label{fig:threehole_rel_error}
        Effect of lag time and number of memory terms on the mean of the absolute value of the relative error for each dynamical statistic  for the two-dimensional triple-well.
    }
\end{figure*}

In this section, we illustrate DGA with memory on a simple model system.
The system is a drift-diffusion process obeying the Fokker--Planck equation
\begin{equation}
\partial_t p_t(x) = \nabla \cdot (p_t(x) \nabla V(x)) + \beta^{-1} \nabla^2 p_t(x)
\end{equation}
on the two-dimensional potential\cite{metzner_illustration_2006}
\begin{equation}
\begin{aligned}[b]
V(x) = {} &
3e^{-\chi_1^2-(\chi_2-1/3)^2} - 3e^{-\chi_1^2-(\chi_2-5/3)^2} \\
& - 5e^{-(\chi_1-1)^2-\chi_2^2} - 5e^{-(\chi_1+1)^2-\chi_2^2} \\
& + 0.2\chi_1^4 + 0.2(\chi_2-1/3)^4,
\end{aligned}
\end{equation}
where $x = (\chi_1,\chi_2)$, as shown in Fig.~\ref{fig:three_pmf}, with inverse temperature $\beta=2$.
We define the reactant state $A$ and product state $B$ as circles with radii $0.25$ centered at $(\pm 1.05, -0.05)$, respectively.

We discretize the system on a $80 \times 80$ grid of equally-spaced points
(with spacing $h=0.05$) in the range $[-2,2] \times [-1.5,2.5]$.
The dynamics of this discretized system follow the master equation \cite{thiede_galerkin_2019,strahan_inexact_2023}
\begin{equation}
\partial_t p_t(x) = \sum_{x'} p_t(x') \gen(x',x),
\end{equation}
where the elements of the matrix $\gen$ are
\begin{equation} \label{eq:3holeL}
\gen(x,x') = \frac{2\beta^{-1}}{h^2} \frac{1}{1+e^{-\beta(V(x)-V(x'))}}
\end{equation}
for $x-x' \in \{(0,\pm h),(\pm h,0)\}$, $\gen(x,x) = -\sum_{x' \ne x} \gen(x,x')$, and $\gen(x,x') = 0$ otherwise.
We use $\gen$ to calculate finite-time operators such as $(\T{t})^\dagger$ and $\ST{B}{t}$. These expressions are given in Appendix~\ref{sec:op_appendix}. We then directly compute the DGA matrices by taking inner products.
For example, we compute \eqref{eq:CtL} as
\begin{equation*}
    \inner{\basis}{(\T{t})^\dagger \basis^\tran} = \sum_{x,x'} \mu(x) \basis(x) (\T{t})^\dagger(x,x') \basis^\tran(x'),
\end{equation*}
where $(\T{t})^\dagger f(x) = \sum_{x'} (\T{t})^\dagger(x,x') f(x')$.

We calculate reference statistics by directly solving \eqref{eq:pi_bvp}, \eqref{eq:mfpt_bvp}, \eqref{eq:qp_bvp}, and \eqref{eq:qm_bvp}.
For the DGA calculations, we use a basis of 64 indicator functions on an equally-spaced $8 \times 8$ grid in the range $[-2,2] \times [-1.5,2.5]$, as shown in Fig.~\ref{fig:three_pmf}b.
Operationally, we first calculate the stationary distribution $\pi$ using \eqref{eq:w_stoc} with an arbitrary distribution of initial conditions and $\gcom = 1$; we subtract the mean from each basis function to normalize $\pi$.
For other statistics, we use the resulting stationary distribution (estimated with the same $\sigma$ and $\tau$) for $\mu$, and we set the value of each basis function to zero for points in $B$ (for $\mfpt$) or $A \cup B$ (for $\qp$ and $\qm$).
We use guess functions $\gmfpt = 0$, $\gqp = \ind{B}$, and $\gqm = \ind{A}$.

In Fig.~\ref{fig:threehole_ref}, we show the stationary distribution ($\pi$), the MFPT to $B$ ($\mfpt$), and the forward and backward committors for the reaction $A$ to $B$ ($\qp$ and $\qm$).
The improvement due to memory is most noticeable in the MFPT, which we plot in Fig.~\ref{fig:threehole_mfpt_compare}. For DGA ($\tau/\sigma = 1$), a lag time of $\tau = 1$ is required for $\imfpt{\sigma,\tau}$ to qualitatively match the reference in Fig.~\ref{fig:threehole_ref}. On the other hand, for DGA with memory ($\tau/\sigma = 5$), $\tau = 0.05$ is sufficient to obtain a reasonable approximation of the reference. When trajectory length is the limiting factor, this shorter lag time directly translates to an order of magnitude savings in simulation time.

We look at the effect of varying parameters $\sigma$ and $\tau$ in Fig.~\ref{fig:threehole_mfpt_error}.  We focus on the  rate constant $k_{AB}$ for the transition from $A$ to $B$ because it is the most sensitive to errors; we calculate it as the inverse of the MFPT from configurations in the $A$ state\cite{reimann_universal_1999}:
\begin{equation}
k_{AB}^{-1} = \int_A \mfpt(x) \pi(x) \odif{x}.
\end{equation}
For this system, the true inverse rate constant is $k_{AB}^{-1} \approx 57$.
In Fig.~\ref{fig:threehole_mfpt_error}a, keeping the number of memory terms constant (by fixing $\tau/\sigma$) and increasing $\tau$ makes the rate constant converge to the correct value.
The best improvement that DGA with memory can give over DGA without memory for a particular $\tau$ (which is limited by the data set) can be seen by comparing the $\tau/\sigma=1$ (DGA without memory) curve with the $\tau/\sigma=10$ curve (which is approximately the same as the $\tau/\sigma=\infty$ limit).
The inverse rate converges at $\tau \approx 10$ without memory and at $\tau \approx 0.05$ with memory.
In Fig.~\ref{fig:threehole_mfpt_error}b, keeping $\sigma$ constant and increasing the number of memory terms also leads to convergence to the correct value.
We note that DGA with memory overestimates the inverse rate around $\tau = 0.05$: memory can overcorrect the error, so one must be careful when assessing convergence since the curve may appear flat at local extrema.
In Fig.~\ref{fig:threehole_mfpt_error}c, we consider the effect of increasing the number of memory terms with fixed $\tau$.
In this case, the result does not converge to the true value but instead retains a fixed error when $\tau$ is too short.
Therefore, to assess convergence, we suggest increasing $\tau$ with either fixed $\tau/\sigma$ or fixed $\sigma$ until the statistic converges.

To demonstrate that these improvements are not specific to the MFPT and the rate constant, we plot the errors in other dynamical statistics in Figs.~\ref{fig:threehole_pw_rel_error} and \ref{fig:threehole_rel_error}.
We calculate relative errors for $\pi$ and $\mfpt$ as $\Delta\isoln{\sigma,\tau}/\soln$, and for $\qp$ and $\qm$ as $\Delta\isoln{\sigma,\tau}/(\soln(1-\soln))$, where $\Delta\isoln{\sigma,\tau}=\isoln{\sigma,\tau}-\soln$.
In Fig.~\ref{fig:threehole_pw_rel_error}, we look at the relative error in each dynamical statistic at $\tau=0.05$.
In all cases, we observe that DGA with memory $(\tau/\sigma=5)$ has a significantly lower error than DGA $(\tau/\sigma=1$).
We note that DGA has large regions of errors with the same sign, while DGA with memory has smaller regions of oscillating errors from memory-corrections with opposing signs.
In Fig.~\ref{fig:threehole_rel_error}, we look at the mean of the absolute value of the relative error in regions with $V \le 0$ (where the system spends the most time; see Fig.~\ref{fig:three_pmf}) for each of the dynamical statistics.
In all cases, DGA with memory outperforms DGA without memory at the same lag time, especially at shorter lag times.

\subsection{\label{sec:aib9} \texorpdfstring{Left-to-right helix transition in AIB\textsubscript{9}}{Left-to-right helix transition in AIB9}}

\begin{figure}[tb]
    \includegraphics{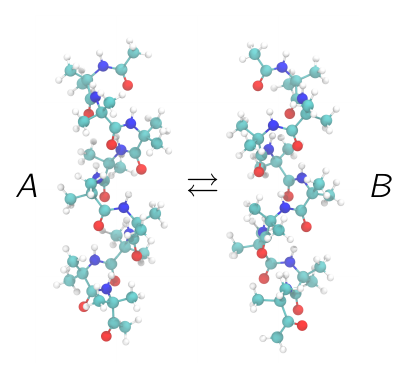}
    \caption{\label{fig:aib_conf}
        $A$ and $B$ configurations of AIB\textsubscript{9}. Carbon, nitrogen, oxygen, and hydrogen atoms are colored cyan, blue, red, and white, respectively.
    }
\end{figure}

\begin{figure}[tb]
    \includegraphics{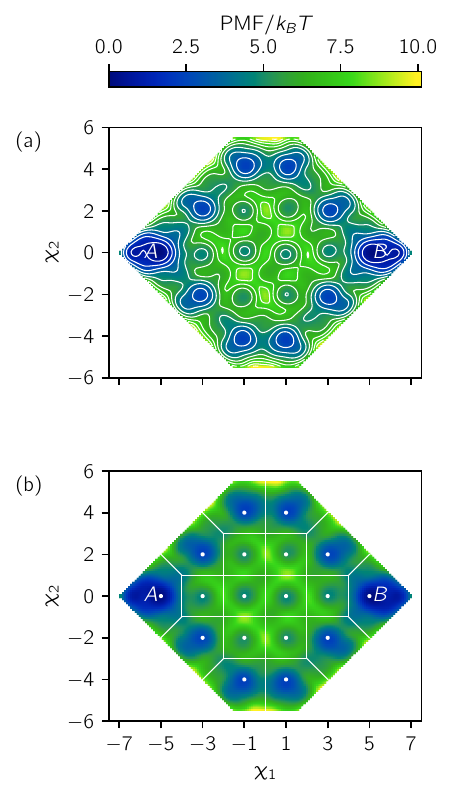}
    \caption{\label{fig:aib_pmf}
        (a) Potential of mean force of AIB\textsubscript{9} in the ($\chi_1,\chi_2)$ CV space with the 2 metastable states $A$ and $B$ labeled. Contours are drawn every $k_\text{B}T$.
        (b) Location and coordinates of the centers (dots) of the 18 states in the CV space.
        Note that multiple different intermediates (30 different intermediates appear as 16 states) may overlap in the same location in the CV space.
        Edges of the basis functions are indicated by lines.
    }
\end{figure}

\begin{figure*}[tb]
    \includegraphics{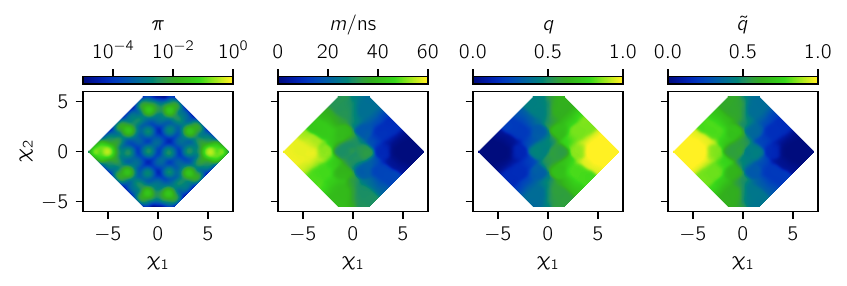}
    \caption{\label{fig:aib_ref}
        Reference statistics for AIB\textsubscript{9} calculated from the long trajectories.
        From left to right, the stationary distribution ($\pi$), the MFPT to the all-right state $B$ ($\mfpt$), the forward committor for the transition from the all-left state $A$ to all-right state $B$ ($\qp$), and the backward committor for the same transition ($\qm$).
    }
\end{figure*}

\begin{figure*}[tb]
    \includegraphics{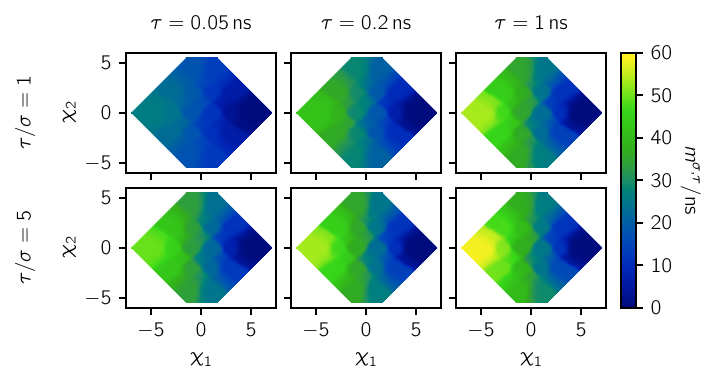}
    \caption{\label{fig:aib_mfpt}
        Mean first passage time to the all-right $B$ state for AIB\textsubscript{9} estimated using DGA without ($\tau/\sigma=1$) and with ($\tau/\sigma=5$) memory, at different lag times $\tau$.
    }
\end{figure*}

\begin{figure*}[tb]
    \includegraphics{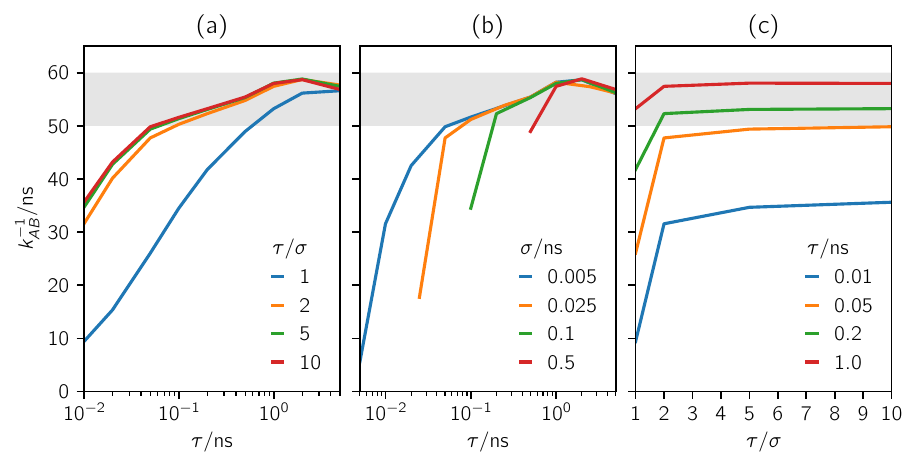}
    \caption{\label{fig:aib_rate}
        Dependence of the estimated AIB\textsubscript{9} inverse rate constant on (a) $\tau$ with fixed $\tau/\sigma$, (b) $\tau$ with fixed $\sigma$, and (c) $\tau/\sigma$ with fixed $\tau$. The true inverse rate constant is between \SI{50}{\nano\second} and $\SI{60}{\nano\second}$, which is indicated by the gray region.
    }
\end{figure*}

\begin{figure*}[tb]
    \includegraphics{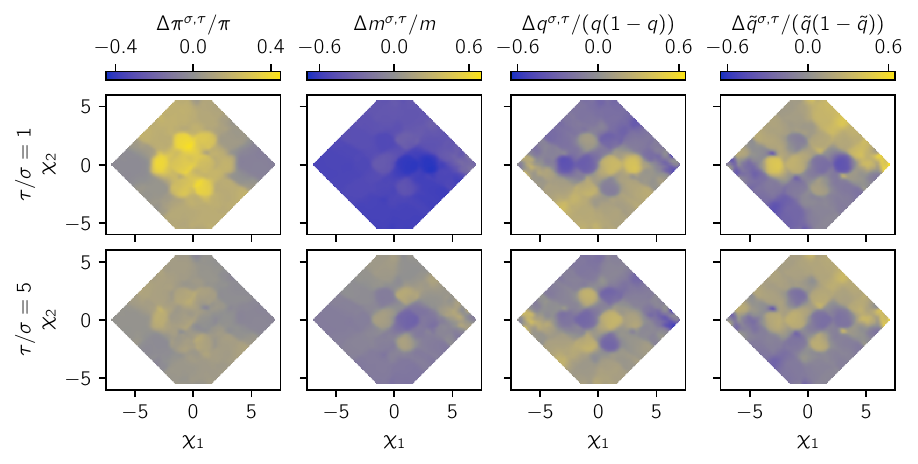}
    \caption{\label{fig:aib_error_compare}
        Relative errors in DGA ($\tau/\sigma=1$) and DGA with memory ($\tau/\sigma=5$) estimates of the AIB\textsubscript{9} dynamical statistics at lag time $\tau = \SI{0.05}{\nano\second}$.
    }
\end{figure*}

\begin{figure*}[tb]
    \includegraphics{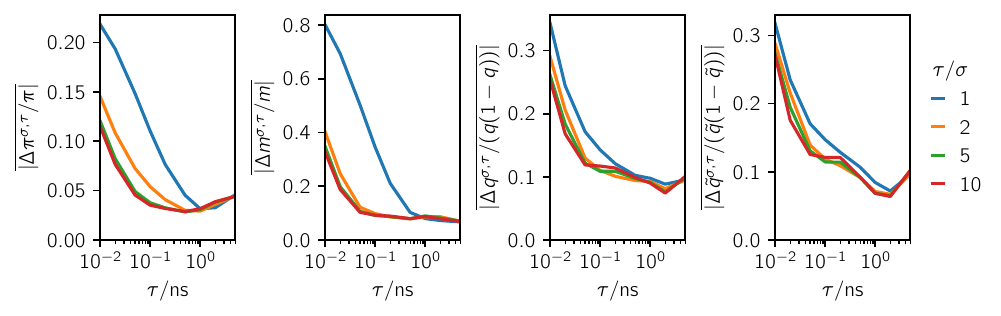}
    \caption{\label{fig:aib_rel_error}
        Effect of lag time and number of memory terms on the mean of the absolute value of the relative error for each dynamical statistic for AIB\textsubscript{9}.
    }
\end{figure*}

To demonstrate DGA with memory for a molecular system, we analyze the dynamics of AIB\textsubscript{9}, a peptide consisting of 9 $\alpha$-aminoisobutyric acid (AIB) residues.
AIB is an unnatural achiral amino acid which forms both left-handed and right-handed 3\textsubscript{10} helices with equal probability.
The left-to-right helix transition was previously studied using MSMs and long, unbiased molecular dynamics simulations \cite{buchenberg_hierarchical_2015, perez_meld-path_2018}.  We recently performed both short and long simulations and used the resulting data to estimate statistics with MSMs and neural networks \cite{strahan_inexact_2023}.

Each amino acid can isomerize between left-handed ($l$) and right-handed ($r$) states, which are defined by the $\phi$ and $\psi$ dihedral angles.
Similarly to Ref.~\onlinecite{strahan_inexact_2023}, we take an amino acid to be in the $l$ state if its dihedral angle values are within a circle of radius $\SI{25}{\degree}$ centered at $(\SI{41}{\degree}, \SI{47}{\degree})$, that is $(\phi - \SI{41}{\degree})^2 + (\psi - \SI{47}{\degree})^2 \le (\SI{25}{\degree})^2$.
Amino acids are classified as being in the $r$ state using the same radius, but centered instead at $(-\SI{41}{\degree}, -\SI{47}{\degree})$.
States $A$ and $B$ are defined by the amino acids at sequence positions 3--7 being all $l$ or all $r$, respectively.
We show such configurations in Fig.~\ref{fig:aib_conf}.
We do not use the two residues on each end of AIB\textsubscript{9} in defining the states as these are typically more flexible \cite{perez_meld-path_2018}.

The short-trajectory data set that we analyze is from Ref.~\onlinecite{strahan_inexact_2023}: it consists of 10 trajectories of duration \SI{20}{\nano\second} from each of the 691 starting configurations in Ref.~\onlinecite{perez_meld-path_2018}.
The data set contains 6,910 trajectories, corresponding to a total sampling time of \SI{138.2}{\micro\second}.
As a reference for comparison, we use 20 simulations of \SI{15}{\micro\second} with the same simulation parameters, corresponding to a total sampling time of \SI{300}{\micro\second}.
In addition, we augment the reference data set with its reflection (since AIB\textsubscript{9} is achiral).

We visualize results on the CV space
\begin{subequations}
\begin{align}
    \chi_1 & = \gamma_3 + \gamma_4 + \gamma_5 + \gamma_6 + \gamma_7, \\
    \chi_2 & = \gamma_3 + \gamma_4 - \gamma_6 - \gamma_7,
\end{align}
\end{subequations}
where $\gamma_i = -0.8 (\sin \phi_i + \sin \psi_i)$.
The $l$ state has $\gamma \approx -1$ and the $r$ state has $\gamma \approx 1$, although these values also contain states other than $l$ and $r$.
We choose these CVs because they can distinguish the 10 most populated states and have physical meaning: $\chi_1$ measures the chirality of a configuration, and $\chi_2$ measures the difference in chirality between the two halves of the molecule.
We project statistics onto the CV space using a kernel density estimate with a Gaussian kernel with a scale parameter of $0.2$.

We compute the potential of mean force (PMF) on the $(\chi_1,\chi_2)$ space from a histogram of CV pairs sampled in the reference data set (Fig.~\ref{fig:aib_pmf}a).
This shows the metastable $A$ and $B$ states on the left and right sides, respectively, as well as the intermediate states connecting them.
In Fig.~\ref{fig:aib_ref}, we show the reference statistics for the stationary distribution, the MFPT to the $B$ state, and the forward and backward committors from the $A$ state to the $B$ state.

For the DGA calculations, we use a basis of 18 indicator functions on the $(\chi_1,\chi_2)$ space as shown in Fig.~\ref{fig:aib_pmf}b.
These indicator functions are Voronoi cells with centers at points $(\chi_1,\chi_2)$ with each $\gamma_i = \pm 1$.
We compute statistics using the same procedure as in Sec.~\ref{sec:threehole}, where we first estimate the stationary distribution and then use it for other statistics. However, here we use short-trajectory data.
We use a rolling window, so all choices of $\sigma$ and $\tau$ use the same amount of data.

In Fig.~\ref{fig:aib_mfpt}, we look at the MFPT, which, as for the two-dimensional triple-well, most clearly shows the improvement due to memory.
At $\tau = \SI{0.05}{\nano\second}$, DGA with memory ($\tau/\sigma=5$) is quite similar to the reference (Fig.~\ref{fig:aib_ref}); DGA without memory ($\tau/\sigma=1$), on the other hand, drastically underestimates the MFPT in all of the CV space.
DGA with memory results at $\tau=\SI{0.2}{\nano\second}$ are comparable to DGA without memory results at $\tau=\SI{1}{\nano\second}$.
Looking at the inverse rate constant from $A$ to $B$ (Fig.~\ref{fig:aib_rate}), which we estimate to be between \SI{50}{\nano\second} and \SI{60}{\nano\second} from the reference trajectories, DGA with memory enters the correct range at $\tau \approx \SI{0.05}{\nano\second}$, whereas DGA without memory requires an order of magnitude more, $\tau \approx \SI{0.5}{\nano\second}$.

We calculate errors by first projecting onto the $(\chi_1,\chi_2)$ space, then using the relative error formulas as described in Sec.~\ref{sec:threehole}.
We restrict the points used to calculate the mean absolute value of the relative absolute error to those with $\lvert\chi_1\rvert + \lvert\chi_2\rvert \le 7$ and $\lvert\chi_2\rvert \le 5.5$ due to a lack of samples outside of this region.
The committors show limited improvement with memory (as shown in Figs.~\ref{fig:aib_error_compare} and \ref{fig:aib_rel_error}), so we focus on the stationary distribution and the MFPT.
In Fig.~\ref{fig:aib_error_compare}, we show the results of DGA ($\tau/\sigma=1$) and DGA with memory ($\tau/\sigma=5$) at $\tau = \SI{0.05}{\nano\second}$.
For the stationary distribution, the states $A$ and $B$ are accurate even at this short lag time for both algorithms, but DGA overestimates the weights of intermediates, especially the minor intermediates in the center.
For the MFPT to the $B$ state, DGA drastically underestimates it in all regions.
Fig.~\ref{fig:aib_rel_error} shows that, for the stationary distribution and MFPT, DGA with memory requires an order magnitude shorter lag time to achieve comparable error to DGA without memory.
Note that, for $\pi$, $\qp$, and $\qm$, the error increases at longer lag times; this is both due to an increase in the impact of sampling error at longer lag times in both the DGA estimates and the reference statistics.
Overall, the relative performances of DGA with and without memory for AIB\textsubscript{9} are qualitatively similar to those for the two-dimensional triple-well, despite the complications that we compute the statistics for AIB\textsubscript{9} from limited simulation data and quantify the error following projection to the $(\chi_1, \chi_2)$ space.

\section{Conclusions}

In this paper we incorporate memory to improve the accuracy of dynamical statistics estimated by dynamical Galerkin approximation.  Our work builds on numerical approaches for solving the orthogonal dynamics equation of the Mori--Zwanzig formalism \cite{darve_computing_2009} and quasi-MSMs \cite{cao_advantages_2020,cao_integrative_2023,dominic_building_2023} and shows that the basic concept behind these approaches  can be generalized and used to compute much more than relaxation time scales.  A key feature of quasi-MSMs and the present approach is that the memory corrections are obtained by directly inverting a GME, so that it is not necessary to assume a particular functional form for the memory.  Quasi-MSMs and DGA with memory also contrast with traditional applications of the Mori--Zwanzig formalism in that the models are high-dimensional.  As a result, the memory decays rapidly, and we find that even a few correction terms are sufficient to achieve high accuracy (with most of the correction coming from the first term beyond the Markov model).  In our numerical experiments, the memory correction decreases the data requirement by an order of magnitude. 

While DGA with memory can decrease the sensitivity of estimates to the choice of basis, a choice must still be made.  To address this issue, we recently introduced an inexact subspace iteration to learn a basis represented by neural networks \cite{strahan_inexact_2023}.  In Ref.~\onlinecite{strahan_inexact_2023}, we apply the memory correction after we learn the basis, but one could apply it after each subspace iteration to accelerate convergence. 
Even when the representation of a dynamical statistic does not have the form in \eqref{eq:dga_expand} (e.g., Ref.~\onlinecite{strahan2023predicting}, or the Richardson iterate in Ref.~\onlinecite{strahan_inexact_2023}), one could define a nonlinear projection operator and in turn compute a memory correction to compensate for deficiencies of the input features.  A separate strategy that can be applied to improve the numerical performance further is to integrate over lag times \cite{lorpaiboon_integrated_2020,cao_integrative_2023}.
It would be interesting to compare the Mori--Zwanzig approach to memory taken here to alternative means of treating memory, such as delay embedding \cite{liebert_proper_1989, das_delay-coordinate_2019, kamb2020time, thiede_galerkin_2019, strahan_long-time-scale_2021}, with regard to data requirements and robustness to hyperparameter choices for high-dimensional models like those considered here.

\begin{acknowledgments}
We are grateful to Xuhui Huang for making us aware of Refs.~\onlinecite{cao_advantages_2020,cao_integrative_2023,dominic_building_2023}.
This work was supported by National Institutes of Health award R35 GM136381 and National Science Foundation award DMS-2054306.
S.C.G.\ acknowledges support by the National Science Foundation Graduate Research Fellowship under Grant No.~2140001.
This work was completed in part with resources provided by the University of Chicago Research Computing Center and we are grateful for their assistance with the calculations.
 ``Beagle-3: A Shared GPU Cluster for Biomolecular Sciences'' is supported by the National Institute of Health (NIH) under the High-End Instrumentation (HEI) grant program award 1S10OD028655-0.
\end{acknowledgments}

\bibliography{ref}

\appendix
\section{\label{sec:iterappendix} Graphical depiction of the effect of memory}

\begin{figure*}[tb]
    \includegraphics{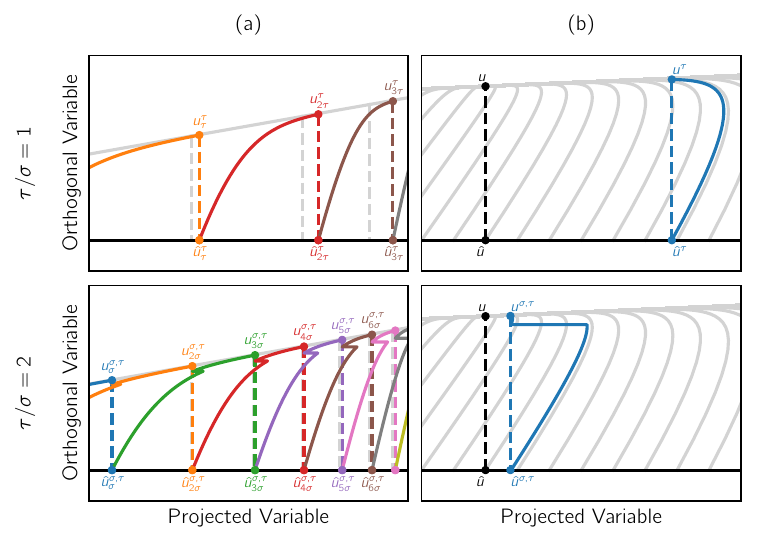}
    \caption{\label{fig:iteration_diagram}
        Schematic of the iterations and their solutions.
        (a) Inexact Richardson iteration corresponding to DGA ($\tau/\sigma=1$, \eqref{eq:inexact}) and DGA with memory ($\tau/\sigma=2$, \eqref{eq:inexactm}). The exact Richardson iteration in \eqref{eq:exact} is plotted is gray.
        (b) Solutions to DGA ($\tau/\sigma=1$) and DGA with memory ($\tau/\sigma=2$). To guide the eye, exact Richardson iterations starting at different projections are plotted in gray.
    }
\end{figure*}

We compare DGA without ($\tau/\sigma=1$) and with ($\tau/\sigma=2$) memory graphically in Fig.~\ref{fig:iteration_diagram}.
The projected dynamics move the system horizontally, while the orthogonal dynamics move it vertically.  We show early iterations in Fig.~\ref{fig:iteration_diagram}a and the fixed point in Fig.~\ref{fig:iteration_diagram}b.
The iterations alternate between using length-$\tau$ trajectories to propagate the projected function $\ipstep{\tau}{t}$ to $\istep{\tau}{t+\tau}$ (colored curves) and projecting back onto the subspace spanned by the basis (dashed lines).  

In the case of DGA without memory (Fig.~\ref{fig:iteration_diagram}a, $\tau/\sigma=1$), the iteration \eqref{eq:inexact} rapidly diverges from the exact iteration in \eqref{eq:exact}, which is plotted in gray.
The fixed point \eqref{eq:inexact} occurs when propagating $\ipsoln{\tau}$ for lag time $\tau$ to $\isoln{\tau}$ such that projecting returns the same $\ipsoln{\tau}$ (Fig.~\ref{fig:iteration_diagram}b).
As drawn, fast variables propagate $\ipsoln{\tau}$ up and to the right, and slow variables ultimately return the system so as to be directly above $\ipsoln{\tau}$, yielding the blue curve.
In the absence of error due to sampling trajectories, $\ipsoln{\tau}$ deviates from $\psoln$ because $\tau$ is insufficiently long for $\isoln{\tau}$ to reach $\soln$.

Comparing the iteration without memory ($\tau/\sigma=1$) to the iteration with memory ($\tau/\sigma=2$), we see that the latter approximates the exact iteration with significantly greater accuracy. Each iteration with memory shown in the figure consists of four steps: (1) propagating $\ipstep{\sigma,\tau}{t}$ by time $\sigma$ (which initially moves the system up and to the right as drawn); (2) adjusting the projected variable with the orthogonal projection so that the projection is $\ipstep{\sigma,\tau}{\sigma}$  (which moves the system to the left horizontally); (3) propagating by another time $\sigma$ to $\istep{\sigma,\tau}{t+2\sigma}$ (which again moves the system up and to the right); (4) projecting to $\ipstep{\sigma,\tau}{t+2\sigma}$ (which moves the system vertically).
More generally, $\tau/\sigma$ steps, each of which propagates the system for time $\sigma$, and $(\tau/\sigma)-1$ projected value adjustments are performed in each iteration, and $(\tau/\sigma)-1$ iterations are inserted between those corresponding to DGA without memory.

As for the case of DGA without memory, the fixed point is reached when the final projection in each step of the iteration returns the system to $\ipsoln{\sigma,\tau}$ (Fig.~\ref{fig:iteration_diagram}b, bottom).
In DGA without memory, the fast modes propagate $\ipsoln{\tau}$ toward the upper right; $\tau$ must be long enough for the slow mode to correct this additional rightward shift.
The impact of the memory correction is to eliminate this shift, so that if $\tau$ is sufficiently large to relax the fast modes that affect the projected variable, the deviation from $\soln$ depends only on the interaction between the slow modes and the orthogonal variables.

\section{\label{sec:op_appendix} Computing finite-time operators using the infinitesimal generator}

In this section, we express the finite-time operators (e.g., $\T{t}$ and $\ST{B}{t}$) in terms of the infinitesimal generator
\begin{equation}
    \gen = \lim_{\dt \to 0^+} \frac{\T{\dt} - \I}{\dt}.
\end{equation}
For a continuous-time Markov jump process on a finite discrete state space, $\gen$ is a matrix (e.g., \eqref{eq:3holeL}) and functions on the grid are vectors, so 
\begin{equation}
    \gen f(x) = \sum_{x'} \gen(x,x') f(x'),
\end{equation}
and we can evaluate expressions in terms of $\gen$ directly. 

\subsection{\label{sec:op_appendix_pi} Transition operator}

The transition operator can be expressed as
\begin{equation}
    \T{t} = (\T{\dt})^{t/\dt} = \lim_{\dt \to 0^+} (\I + \dt \gen)^{t/\dt} = e^{t \gen}.
\end{equation}
Similarly,
\begin{equation}
    (\T{t})^\dagger(x,x') = e^{t \gen^\dagger}(x,x'),
\end{equation}
and
\begin{equation}
    \gen^\dagger(x,x') = \mu(x') \gen(x',x) / \mu(x)
\end{equation}
from the definition of the adjoint.

\begin{widetext}

\subsection{\label{sec:op_appendix_p} Stopped transition operator}

Here, we express the stopped transition operator in terms of the infinitesimal generator. To this end, we observe that
\begin{equation}\label{eq:fwdstopped}
    f(\X{t\bmin\stp{B}})
        = \lim_{\dt \to 0^+} \biggl(f(\X{0}) + \sum_{n=0}^{(t/\dt)-1} \biggl[\prod_{n'=0}^n \ind{B^\comp}(\X{n'\dt})\biggr] (f(\X{(n+1)\dt}) - f(\X{n\dt}))\biggr).
\end{equation}
That is, at time $n\dt$, we propagate the process forward-in-time only if $n\dt < \stp{B}$ (the product above is 1 only if $\X{0},\ldots,\X{n\dt} \in B^\comp$ and 0 otherwise).
Therefore, we can express \eqref{eq:Sp} as
\begin{align}
    \ST{B}{t} f(x)
    & = \E[f(\X{t\bmin\stp{B}}) \mid \X{0} = x] \label{eq:Sp_exp} \\
    & = \lim_{\dt \to 0^+} \biggl(f(x) + \sum_{n=0}^{(t/\dt)-1} (\D{\ind{B^\comp}} \T{\dt})^n \D{\ind{B^\comp}} (\T{\dt} - \I) f(x)\biggr)
    \label{eq:Sp_step2_2} \\
    & = f(x) + \int_0^t e^{t' \D{\ind{B^\comp}} \gen} \D{\ind{B^\comp}} \gen f(x) \odif{t'}
    \label{eq:Sp_step3} \\
    & = e^{t \D{\ind{B^\comp}} \gen} f(x).
    \label{eq:Sp_step4}
\end{align}
We substitute \eqref{eq:fwdstopped} into \eqref{eq:Sp_exp} and express the resulting conditional expectation in terms of $\T{\dt}(x,x') = \Pr[\X{t+\dt}=x' \mid \X{t}=x]$ to obtain \eqref{eq:Sp_step2_2}.
Taking the limit $\dt \to 0^+$ yields \eqref{eq:Sp_step3}, and evaluating the integral yields \eqref{eq:Sp_step4}.
We define $\D{g}$ to be a diagonal matrix with entries $\D{g}(x,x) = g(x)$;
the matrix $\D{\ind{B^\comp}} \gen$ has entries
\begin{equation}
(\D{\ind{B^\comp}} \gen)(x,x') = \ind{B^\comp}(x) \gen(x,x').
\end{equation}
We evaluate the integral $\int_0^t \ST{B}{t'} \ind{B^\comp} \odif{t'} = \int_0^t e^{t' \D{\ind{B^\comp}} \gen} \ind{B^\comp} \odif{t'}$ in \eqref{eq:mfpt_h} using the identity
\begin{align}\label{eq:matrixexp}
    \exp\left(t
        \begin{bmatrix}
            \D{\ind{B^\comp}} \gen & \ind{B^\comp} \\ 0 & 0
        \end{bmatrix}
    \right)
    & =
    \lim_{\dt \to 0^+}
    \left(
        \begin{bmatrix}
            \I & 0 \\ 0 & \I
        \end{bmatrix}
        + \dt
        \begin{bmatrix}
            \D{\ind{B^\comp}} \gen & \ind{B^\comp} \\ 0 & 0
        \end{bmatrix}
    \right)^{t/\dt}
    \nonumber \\
    & =
    \lim_{\dt \to 0^+}
    \begin{bmatrix}
        (\I + \dt \D{\ind{B^\comp}} \gen)^{t/\dt} & \dt \sum_{n=0}^{(t/\dt)-1} (\I + \dt \D{\ind{B^\comp}} \gen)^n \ind{B^\comp} \\ 0 & \I
    \end{bmatrix}
    \nonumber \\
    & =
    \begin{bmatrix}
        e^{t \D{\ind{B^\comp}} \gen} & \int_0^t e^{t' \D{\ind{B^\comp}} \gen} \ind{B^\comp} \odif{t'} \\ 0 & 1
    \end{bmatrix},
\end{align}
That is, we compute the exponential and then take only the upper right element of the result.

\subsection{\label{sec:op_appendix_qm} Stopped transition operator for the time-reversed process}

Similarly to \eqref{eq:fwdstopped} to \eqref{eq:Sp_step4}, we use the identity
\begin{equation} \label{eq:bwdstopped}
    f(\X{-(t\bmin\stm{A \cup B})})
        = \lim_{\dt \to 0^+} \biggl(f(\X{0}) + \sum_{n=0}^{(t/\dt)-1} \biggl[\prod_{n'=0}^n \ind{(A \cup B)^\comp}(\X{-n'\dt})\biggr] (f(\X{-(n+1)\dt}) - f(\X{-n\dt}))
        \biggr)
\end{equation}
to express \eqref{eq:g_Sm_f} as
\begin{align}
\inner{g}{\ST{A \cup B}{-t} f}
& = \E\biggl[g(\X{0}) f(\X{-(t\bmin\stm{A \cup B})}) \frac{\com(\X{-\tau})}{\com(\X{0})} \biggm| \X{-\tau} \sim \mu\biggr]
\label{eq:Sm_exp} \\
&
\begin{aligned}[b]
    {} =
    \lim_{\dt \to 0^+}
        \sum_x
        \frac{g(x) \mu(x)}{\com(x)}
        \biggl(
            f(x) (\T{\tau})^\dagger \com(x)
            + \sum_{n=0}^{(t/\dt)-1} \biggl[
                &
                (\D{\ind{(A \cup B)^\comp}} (\T{\dt})^\dagger)^n
                \D{\ind{(A \cup B)^\comp}}
                \\
                & \times
                ((\T{\dt})^\dagger \D{f} - \D{f} (\T{\dt})^\dagger)
                (\T{\tau-(n+1)\dt})^\dagger \com(x)
                \biggr]
        \biggr)
\end{aligned}
\label{eq:Sm_step2_2} \\
& = \sum_x \frac{g(x) \mu(x)}{\com(x)} \biggl(f(x) e^{\tau \gen^\dagger} \com(x) + \int_0^t e^{t' \D{\ind{(A \cup B)^\comp}} \gen^\dagger} \D{\ind{(A \cup B)^\comp}} (\gen^\dagger \D{f} - \D{f} \gen^\dagger) e^{(\tau-t') \gen^\dagger} \com(x) \odif{t'}\biggr).
\label{eq:Sm_inner}
\end{align}
We substitute \eqref{eq:bwdstopped} into \eqref{eq:Sm_exp} and expand the resulting expectation in terms of $(\T{\dt})^\dagger(x,x') = \Pr[\X{t+\dt}=x \mid \X{t}=x'] \mu(x') / \mu(x)$ to obtain \eqref{eq:Sm_step2_2}.
Taking the limit $\dt \to 0^+$ yields \eqref{eq:Sm_inner}, where the entries of the matrices $\D{\ind{(A \cup B)^\comp}} \gen^\dagger$ and $\D{\ind{(A \cup B)^\comp}} (\gen^\dagger \D{f} - \D{f} \gen^\dagger)$ are
\begin{align}
(\D{\ind{(A \cup B)^\comp}} \gen^\dagger)(x,x') & = \ind{(A \cup B)^\comp}(x) \gen^\dagger(x,x'), \\
(\D{\ind{(A \cup B)^\comp}} (\gen^\dagger \D{f} - \D{f} \gen^\dagger))(x,x') & = \ind{(A \cup B)^\comp}(x) \gen^\dagger(x,x') (f(x') - f(x)).
\end{align}
Likewise,
\begin{align}
    \inner{g}{(\ST{A \cup B}{-t}-\I) f}
    & = \E\biggl[g(\X{0}) (f(\X{-(t\bmin\stm{A \cup B})}) - f(\X{0})) \frac{\com(\X{-\tau})}{\com(\X{0})} \biggm| \X{-\tau} \sim \mu\biggr]
    \label{eq:SmI_exp} \\
    & = \sum_x \frac{g(x) \mu(x)}{\com(x)} \int_0^t e^{t' \D{\ind{(A \cup B)^\comp}} \gen^\dagger} \D{\ind{(A \cup B)^\comp}} (\gen^\dagger \D{f} - \D{f} \gen^\dagger) e^{(\tau-t') \gen^\dagger} \com(x) \odif{t'}.
    \label{eq:SmI_inner}
\end{align}
The integrals in \eqref{eq:Sm_inner} and \eqref{eq:SmI_inner} can be evaluated similarly to \eqref{eq:matrixexp}:
\begin{align}
    & \exp\left(t
    \begin{bmatrix}
        \D{\ind{(A \cup B)^\comp}} \gen^\dagger & \D{\ind{(A \cup B)^\comp}} (\gen^\dagger \D{f} - \D{f} \gen^\dagger) \\ 0 & \gen^\dagger
    \end{bmatrix}
    \right) \\
    & \quad {} =
    \begin{bmatrix}
        e^{t \D{\ind{(A \cup B)^\comp}} \gen^\dagger} & \int_0^t e^{t' \D{\ind{(A \cup B)^\comp}} \gen^\dagger} \D{\ind{(A \cup B)^\comp}} (\gen^\dagger \D{f} - \D{f} \gen^\dagger) e^{(t-t') \gen^\dagger} \odif{t'} \\ 0 & e^{\gen^\dagger t}
    \end{bmatrix}.
\end{align}
The expressions for $\ST{A \cup B}{-t}$ are more complicated than those for $\ST{A \cup B}{t}$ because we want \eqref{eq:Sm_inner} and \eqref{eq:SmI_inner} to be equivalent to \eqref{eq:Sm_exp} and \eqref{eq:SmI_exp}, respectively, when $\com$ is approximated. If we evaluate them without approximating $\com$, $\gen^\dagger \com = 0$ and $e^{\gen^\dagger t} \com = \com$; therefore, $\inner{g}{\ST{A \cup B}{-t} f} = \sum_{x,x'} (g(x) \mu(x) / \com(x)) e^{t \D{\ind{(A \cup B)^\comp}} \gen^\dagger}(x,x') f(x') \com(x')$ and $\inner{g}{(\ST{A \cup B}{-t}-\I) f} = \sum_{x,x'} (g(x) \mu(x) / \com(x)) (e^{t \D{\ind{(A \cup B)^\comp}} \gen^\dagger}-\I)(x,x') f(x') \com(x')$.

\end{widetext}
~
\end{document}